\documentclass[journal=aamick,manuscript=article,layout=onecolumn,email]{achemso}
\usepackage{natbib}
\usepackage[version=3]{mhchem} 
\PassOptionsToPackage{linktocpage}{hyperref}
\usepackage{bm}
\usepackage{amssymb}
\usepackage{amsfonts}
\usepackage{amsmath}
\usepackage{graphicx}
\usepackage{dcolumn}
\usepackage{graphicx}
\usepackage{multirow}
\usepackage{setspace}
\usepackage{longtable}

\author{Thaneshwor P. Kaloni}
\affiliation{Department of Chemistry, University of Manitoba, Winnipeg, MB, R3T 2N2, Canada}
\email{thaneshwor.kaloni@umanitoba.ca, +1-204-952-2900}
\author{Mohsen Modarresi}
\affiliation{Department of Physics, Ferdowsi University of Mashhad, Mashhad, Iran}
\author{Muhammad Tahir}
\affiliation{Department of Physics, Concordia University, 7141 Sherbrooke Ouest Montr\'eal, Canada H4B 1R6}
\author{Mahmood Rezaee Roknabadi}
\affiliation{Department of Physics, Ferdowsi University of Mashhad, Mashhad, Iran}
\author{Georg Schreckenbach}
\affiliation{Department of Chemistry, University of Manitoba, Winnipeg, MB, R3T 2N2, Canada}
\author{Michael S. Freund}
\affiliation{Department of Chemistry, University of Manitoba, Winnipeg, MB, R3T 2N2, Canada}

\title{Electrically Engineered Band Gap in Two-Dimensional Ge, Sn, and Pb: A First-Principles and Tight-Binding Approach}

\begin{document}
\newpage

\begin{abstract}
First-principles calculations were performed to investigate the electronic structure of two-dimensional (2-D) Ge, Sn, and Pb without and with the presence of an external electric field in combination with spin-orbit coupling. Tight-binding calculations based on four orbitals per atom and an effective single orbital are presented to match with the results obtained from first-principles calculations. In particular, the electronic band structure and the band splitting are investigated with both models. Moreover, the simple $k\cdot p$ model is also considered in order to understand the band splitting in the presence of an external electric field and spin-orbit coupling. A large splitting is obtained, which is expected to be useful for spintronic devices. The fair agreement between the first-principle, $k\cdot p$ model, and tight-binding approaches leads to a table of parameters for future tight-binding studies on hexagonal 2-D nanostructures. By using the tight binding parameters, the transport properties of typical 0-D triangular quantum dots between two semi-infinite electrodes in the presence of spin-orbit coupling are addressed.
\end{abstract}

\newpage

\section{INTRODUCTION}
After the discovery of graphene, many two-dimensional (2-D) materials have been predicted, in particular, materials composed of group IV elements.\cite{Novoselov22102004,nn400280c,geim2013vander,rsc-amin} These materials are very interesting not only for fundamental research but also for electronic device applications.\cite{PhysRevLett.102.236804} For example, H-passivated graphene called graphane \cite{Elias30012009,PhysRevLett.111.136804}, a silicon analog of graphene called silicene \cite{PhysRevLett.108.155501}, and a germanium counterpart of graphene called germanene \cite{nn4009406} have attracted the interest of many researchers. Silicene and graphene are relatively similar in nature due to the fact that Si and C belong to the same group IV in the periodic table. The $sp^3$ hybridization is energetically favourable in case of Si, while $sp^2$ hybridization is energetically favourable for graphene. Experimentally, silicene has been synthesized on metallic substrates, such as Ag \cite{padova,doi:10.1021/nl072591y,Quaresima}, ZrB$_2$ \cite{PhysRevLett.108.245501}, 
and Ir(111).\cite{doi:10.1021/nl304347w} Moreover, theoretically it has also been studied on semiconducting substrates, such as $h$-BN and SiC \cite{kaloni2013quasifreestanding,PhysRevB.88.235418,doi:10.1021/jp311836m} and it has been found that the semiconducting substrates could potentially be a way for obtaining free-standing silicene. First-principles calculations have shown that silicene is particularly interesting due to the formation of graphene-like dispersion.\cite{PhysRevB.79.115409,kaloni-jap,PhysRevB.87.085423,PhysRevB.89.035409,PSSR:PSSR201409245,doi:10.1021/jp505814v}

A rather weak interaction of germanene with the semiconducting GaAs(0001) substrate has been demonstrated in such a way that a quasi-free-standing germanene could be realized.\cite{kaloni2013-germanene} Without H passivation of the dangling bonds of Ga terminated GaAs(0001), the support strongly interacts with the substrate, while the interaction can be strongly reduced by H intercalation at the interface of GaAs(0001) and germanene. Recently, germanene has been synthesized on a Pt(111) surface. \cite{ADMA:ADMA201400909,Bampoulis} The grown germanene exhibits a buckled structure where a $3\times3\times1$ supercell of germanene coincides with a $\sqrt19\times\sqrt19\times1$ supercell of the Pt(111) surface. The structural, electronic, and magnetic properties of germanene can be influenced strongly by doping. For example, the Quantum Anomalous Hall effect has been predicted for V-doped germanene\cite{kaloni-accept-jpcc} and the Quantum Spin Hall effect \cite{PhysRevB.89.195403} has been predicted for H-doped germanene. Furthermore, the application of mechanical strain also influences greatly the structural properties of germanene.\cite{Kaloni2013137,Wang20136}

Since the ionic radii of the elements increase from Ge to Pb, they increasingly promote $sp^3$ hybridization as a result. The magnitude of the buckling in 2-D C, Si, Ge, Sn, Pb is found to be 0.00 \AA, 0.46 \AA, 0.68 \AA, 0.84 \AA, and 0.89 \AA, respectively.\cite{tsai2013gatedsilicene} In fact, the buckling plays a crucial role in the engineering of the electronic structure (band gap) in these materials.\cite{Ni,kaloni2013quasifreestanding,doi:10.1021/jp505814v,msf1} Spin-orbit coupling is strong in these novel 2-D materials, especially for the heavier atoms. Thus, it is important to incorporate it into the theoretical studies in order to achieve accurate results. Materials with strong spin-orbit coupling are potential candidates for applications in spintronic devices \cite{prl2011}. As already mentioned, these materials are buckled and they produces a very small band gap. The buckled structure is comparatively favorable for the application of a perpendicular electric field, which in fact allows tuning the band gap 
easily.\cite{Ni} The band gap, in turn, is essential to operate electronic devices such as transistors. Recently, silicene-based transistors have been realized experimentally, which can be operated at room temperature.\cite{natnatech2015} Thus we are expecting interesting results for the systems under study. 

The magnitude of the spin-orbit coupling is found to be 1.$10^{-3}$ meV, 4.2 meV, 11.8 meV, 36.0 meV, and 207.3 meV, respectively for 2-D C, Si, Ge, Sn, Pb.\cite{tsai2013gatedsilicene} The combined effects of the electric field and spin-orbit coupling for 2-D Ge, Sn, and Pb have not been addressed so far. To addressed this issue, density functional theory (DFT) based calculations have been performed and the obtained results were verified with the help of the tight-binding (TB) approach.\cite{TB} In particular, the electronic structure (the nature of the band gap) and the band splitting were investigated. Further, the $k\cdot p$ model \cite{KP} was used to understand the band splitting. Excellent agreement between the DFT and TB calculations was found. The obtained TB parameters can therefore be used for the investigation of larger nanostructures. As an example and proof of principle, we address the effects of spin-orbit coupling on the transport through a 0-D triangular quantum dot (TQD) between two semi-infinite electrodes.

\section{METHODOLOGY}
\subsection{DENSITY FUNCTIONAL THEORY (DFT) CALCULATIONS}
DFT calculations have been performed with the generalized gradient approximation (PBE) \cite{PBE} as implemented in the Vienna Ab-initio code (VASP).\cite{VASP} A plane wave cutoff energy of 450 eV and a Monkhorst-Pack k-grid of $24\times24\times1$ have been employed.\cite{MONKHORST} A $1\times1\times1$ unit cell of 2-D Ge, Sn, and Pb with a lattice constant $a$ of 4.06 \AA, 4.67 \AA, and 4.93 \AA\, respectively, has been taken from Ref.\cite{tsai2013gatedsilicene} Moreover, a finite external electric field has been applied by using the approach described in Refs.\cite{PhysRevB.51.4014,PhysRevB.46.16067}

\subsection{TIGHT-BINDING APPROACH} 
To verify the results obtained from the DFT calculations, a TB Hamiltonian has been used for a unit cell that contains two 
atoms by following Ref.\ \cite{PhysRevB.78.205425} Only the hopping between nearest neighbors in the unit cell was considered. The $sp^3$ basis has been employed to expand the Hamiltonian. As a result the total Hamiltonian of the unit cell becomes a $8\times8$ matrix, which is given by 
\begin{equation}
H_0=\sum_{i,\alpha,\sigma}\epsilon_{i,\alpha}C_{i,\alpha,\sigma}^\dagger C_{i,\alpha,\sigma} + \sum_{<i,j>,\alpha,\beta,\sigma}t_{i,j,\alpha,\beta}C_{i,\alpha,\sigma}^\dagger C_{i,\alpha,\sigma}.
\end{equation}  
Here, $C_{i,\alpha,\sigma}^\dagger$ and $C_{i,\alpha,\sigma}$ represent the electron creation and annihilation operators in the hexagonal lattice sites for electrons with spin ($\sigma$) and orbital ($\alpha$), respectively and $<i,j>$ represents pair 
of nearest neighbor. The first term of eq$^n$ (1) is the on-site energy for $\alpha$ of the i$^{th}$ atom, and the second term describes the hopping between nearest neighbor atomic sites $<i,j>$ for different values of $\alpha$. The hopping parameters were adopted from the Slater-Koster method.\cite{PhysRev.94.1498}

The Hamiltonian in the reciprocal space is obtained by performing a Fourier transformation of the real-space Hamiltonian. The band structure is given by the k-dependent eigenstates of the transformed Hamiltonian matrix. Due to the buckling in the atomic structure of the hexagonal lattices, the presence of a static external electric field induces a voltage difference between the two atomic positions in the unit cell. Assuming that the electric field is homogeneous along the $z$-direction, the electric field included Hamiltonian can thus be written as, $H^E=E\sum_{i,\alpha,\sigma}z_iC_{i,\alpha,\sigma}^\dagger C_{i,\alpha,\sigma}$, where $z_i$ is the $z$-component of the atomic position of the atom ``\textit{i}". Therefore, the effect of an external electric field is similar to a site dependent on-site energy for each atom. In principle, the buckling of the hexagonal lattices enhances the overlap between $\pi$ and $\sigma$ states, and thus, stabilizes the system. As a result, one should consider at least four orbitals per atom to study these nanostructures. 

The interaction between electron spin and orbital angular momentum couples both degrees of freedom to each other. The spin-orbit coupled Hamiltonian can be expressed as,

\begin{equation}
H^{SOC}=\lambda\vec{L}.\vec{S}=\lambda\left[\frac{1}{2}(L_{+}S_{-}+L_{-}S_{+})+L_zS_z\right],
\end{equation}
where $\lambda$ is the strength of the spin-orbit coupling, and $L_{\pm}$ and $S_{\pm}$ are the ladder operators for orbital and spin angular momentum, respectively.\cite{PhysRevB.82.245412} In the presence of both spin-orbit coupling and an external electric field, the total Hamiltonian becomes $H^{tot}=H_0+H^E+H^{SOC}$. By inclusion of the spin degrees of freedom, the total Hamiltonian becomes a $16\times16$ matrix. For the sake of simplicity, the overlap matrix is taken as an identity matrix and the orbital dependence of the strength of the spin-orbit coupling is ignored. 

An effective Kene and Mele Hamiltonian is also considered to describe the 
hexagonal lattices.\cite{PhysRevLett.95.226801,PhysRevLett.95.146802} It can be expressed as, 

\begin{equation}
H^{eff}=\sum_{<i,j>}t_{i,j}^{eff}C_{i}^\dagger C_{j}+i\lambda^{eff}\sum_{<<i,j>>}V_{ij}C_{i}^\dagger s^z C_{j}.
\end{equation}

Here, s$_z$ is the $z$-component of the Pauli matrix and $t_{i,j}^{eff}$ and $\lambda^{eff}$ are the effective hopping between nearest neighbor sites and spin-orbit strength for next nearest neighbor sites, respectively. while $V_{ij}=\pm1$, $+/-$ 
represents  clockwise/anti-clockwise orientation of the hybrid orbitals at the two nearest neighbor sites i and j. 

By using the effective model the electric transport is studied through a TQD between two semi-infinite structure-less electrodes as shown in Fig.\ 1. The TQD has localized zero-energy states on the edges, which are split up or down in the presence of a Hubbard term by inducing magnetic moments on the zigzag edges.\cite{PhysRevB.83.174441,modarresi2014studyof,modarresi2014spindependent,PhysRevB.85.075431} The Green’s function approach \cite{GF} is a well-known method for calculating the transport properties through a nano-structure between two leads. Here, a TQD between electrodes has been studied in the presence of intrinsic spin-orbit coupling.

\begin{figure}[t!]
\includegraphics[width=0.40\textwidth,clip]{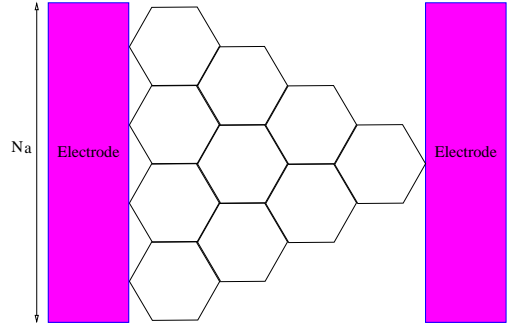}
\caption{Symbolic representation of the top view of a TQD between two semi-infinite electrodes, where N$_{a}$ is the number of atoms on the zigzag edge, which determines the size of TQD.}
\end{figure}

In order to describe the above system, one should consider a Hamiltonian that contains the TQD, the left and right electrodes and the coupling between them. The net effect of the two semi-infinite electrodes is described by using self-energy terms that can be calculated by a simple iterative method. Here, for simplicity the wide-band approximation is used which ignores the real part of the self-energy and estimates the imaginary part with a constant value for all energy intervals.\cite{modarresi2014spinpolarization} The self-energy operator is useful in calculating the retarded Green’s function, defined as, $G^r(\epsilon)=\left((\epsilon+i\delta^+)I-H_{TQD}-\sum_L-\sum_R\right)^{-1}$. In this equation, $\sum_{L/R}$ are the self-energy operators for the left and right electrodes, $I$ is the identity matrix and $H_{TQD}$ is the Hamiltonian of the quantum dot. The transmission spectrum as a function of energy through the quantum dot is calculated by using the Fisher-Lee relation \cite{FL} $T(\epsilon)=Trace\left[\Gamma_LG^r(\epsilon)\Gamma_RG^r(\epsilon)\right]$, where $\Gamma_{L/R}=2i\sum_{L/R}^r$ is the conductance of the noninteracting electrons. The current-voltage curve is calculated by using the Landauer-Buttiker equation and the transmission spectrum.\cite{LB} 

\section{RESULTS AND DISCUSSION}
\subsection{ELECTRONIC STRUCTURE}
\begin{table*}[h]
\begin{tabular}{|c|c|c|c|c|c|c|c|c|c|c|c|c|}
\hline
System& \multicolumn{4}{|c|}{\multirow{1}{*}{Ge}}&\multicolumn{4}{|c|}{\multirow{1}{*}{Sn}}&\multicolumn{4}{|c|}{\multirow{1}{*}{Pb}}\\
\cline{2-5}
\hline
$E_z$& \multicolumn{2}{|c|}{\multirow{1}{*}{E$_{gap}$}}& \multicolumn{2}{|c|}{\multirow{1}{*}{$\Delta_{sp}$}} &\multicolumn{2}{|c|}{\multirow{1}{*}{E$_{gap}$}}& \multicolumn{2}{|c|}{\multirow{1}{*}{$\Delta_{sp}$}}&\multicolumn{2}{|c|}{\multirow{1}{*}{E$_{gap}$}}& \multicolumn{2}{|c|}{\multirow{1}{*}{$\Delta_{sp}$}}\\
\hline
&DFT&TB&DFT&TB&DFT&TB&DFT&TB&DFT&TB&DFT&TB\\
\hline
0.0&22.9  &34.2 &0.0& 0.0&73.4&77.4 &0.0 &0.0 &409.1&366.3 &0.0 &0.0\\
\hline
0.001&20.1&33.7 &3.5& 0.45 &76.3&76.8 &0.71&0.6 &408.5&365.8 &0.4 &0.5\\
\hline
0.005&23.8&31.5 &23.1&2.7 &78.2& 75.5&0.76&2.9 &406.1&363.8 &4.5 &2.5\\
\hline
0.01&47.4 &58.7 &24.5&5.4 &79.5& 71.3&7.51&6.1 &404.9  &361.3 &5.4 &5.0\\
\hline
0.5&31.5  &242.2 &25.8&35.2 &309.9&227.8 &76.6 & 77.9&157.2&117.1 &266.9 &255.2\\
\hline
1.0&115.4$^*$&518.7$^*$ &26.1 &33.6 &318.3&530.1 &81.2&79.2 &$m$&124.2&0.0&359.2 \\
\hline
1.5&408.8$^*$&792.2$^*$ &34.8&37.7 &328.7&826.9 &86.2&81.3 &$m$&352.4&0.0&357.0\\
\hline
2.0&570.4$^*$&1063$^*$ &43.3 &39.3 &228.3$^*$&1117.2$^*$ & 90.1&84.1 & $m$&564.0&0.0&354.3\\
\hline
\end{tabular}
\caption{The magnitude of the external electric field $E_z$ (in V/\AA), band gap $E_{gap}$ (in meV), and band splitting $\Delta_{sp}$ (in meV) for 2-D Ge, Sn, and Pb, where $m$ represents metallic states and $*$ represents an indirect band gap.} 
\end{table*}

\begin{figure}[t!]
\includegraphics[width=0.6\textwidth,clip]{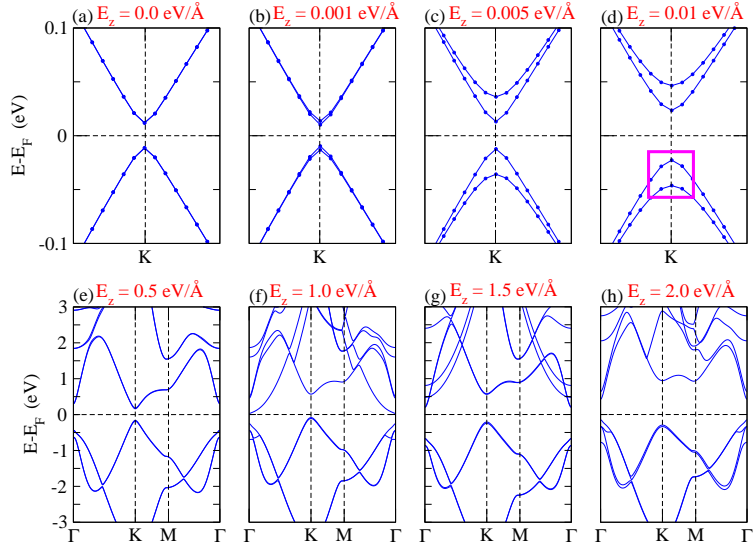}
\caption{The electronic structure of 2-D Ge for different values of the external electric field $E_z$. Only the zoomed area around the K-point is shown for (a-d). The band splitting is represented by a pink square.}
\end{figure}

The electronic structure of 2-D Ge, Sn, and Pb was studied without and with the presence of an external electric field by 
employing DFT calculations and the TB approaches. Spin-orbit coupling has been taken into account in all the calculations. The calculated band structure of 2-D Ge is addressed in Fig.\ 2. The calculated band gap (E$_{gap}$) in the absence of an electric field is found to be 22.9 meV with no band splitting, which is in good agreement with the 
available reports.\cite{kaloni2013-germanene,kaloni-accept-jpcc,PhysRevB.89.195403,Kaloni2013137,Wang20136} The values of E$_{gap}$ and $\Delta_{sp}$ increase with increasing magnitude of the electric field.\cite{Ni} The $\Delta_{sp}$ is increased because of the combined effect of spin-orbit coupling and an external electric field, which agrees well with a previous report for silicene.\cite{PhysRevB.86.195405} Whereas, the band splitting can be attributed to the fact that the external perpendicular electric field removes the space inversion symmetry, the absence of space inversion symmetry lifts the spin degeneracy due to spin-orbit coupling; as a result the splitting between states has been observed.\cite{PhysRevB79} It is worth mentioning that transition from a direct band gap (at the K-point) to an indirect band gap (the valence band maximum and conduction band minimum at the K- and $\Gamma$-points of the Brillouin zone, respectively) is obtained above an external electric field of 1.0 V/\AA. A direct to indirect band gap transition occurs under an external electric field because the electric field enhanced spontaneous polarization. This polarization is responsible for the direct to indirect band gap transition.\cite{C3NR06072A,noborisaka2014electrictuning} In addition, a perpendicular electric field reduces the band gap and thus, the transition is possible.\cite{C4CS00276H,kaloni-jap,Yang2010} The data sets are summarized in Table 1. For low values of the electric field the electronic gap is related to the spin-orbit coupling but for higher values of the external field the gap results from the difference of the electric potential on the different atomic sites. An unscreened uniform electric field has been considered; as a result, for high values of the electric field strength, the tight binding model predicts a relatively bigger electronic gap with respect to the DFT results.

The transition from a direct to an indirect E$_{gap}$ has been observed previously either via strain or via the application of an electric field in 2-D transition-metal dichalcogenides, such as MoS$_2$ and WS$_2$.\cite{rsc-amin,C3NR06072A,new1,new2} In addition, a large $\Delta_{sp}$ value of about 43.3 meV is obtained in 2-D Ge for an electric field of 2.0 V/\AA, which is expected to be useful for spintronic devices.\cite{butler2014,tahir2013valleypolarized} In general, due to the buckled structure of 2-D Ge, the presence of an external electric field induces a difference in the electric potential on the two atoms of the unit cell. The potential opens an energy gap at the K-point which increases with increasing magnitude of the electric field. The spin-orbit coupling together with the electric field splits the up and down states. 

\begin{figure}[ht]
\includegraphics[width=0.5\textwidth,clip]{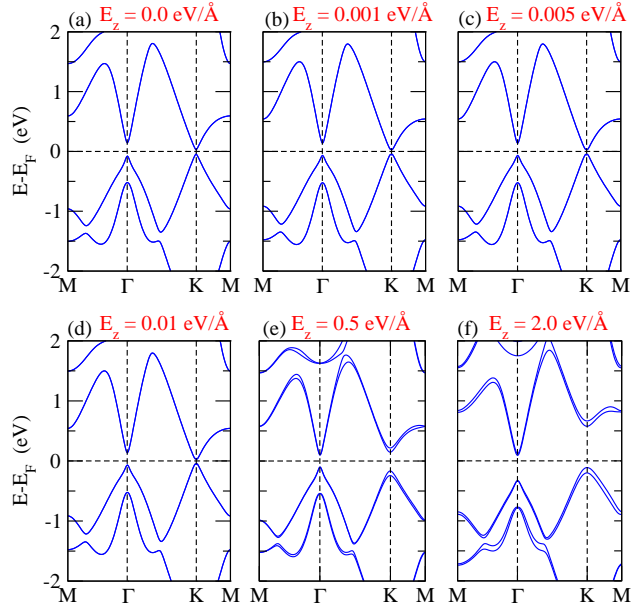}
\caption{The electronic structure of 2-D Sn for different values of the external electric field $E_z$}
\end{figure}

The calculated electronic band structure of 2-D Sn is shown in Fig.\ 3. A E$_{gap}$ of 73.4 meV with $\Delta_{sp}=0$ is obtained in the absence of an external electric field, see Fig.\ 3(a). It has been reported that 2-D Sn also forms a Dirac-like cone without inclusion of the spin-orbit coupling in the calculations.\cite{PhysRevLett.111.136804} The authors of Ref.\ \cite{PhysRevLett.111.136804} observed a band gap of about 100 meV by removing the degeneracy between the bands at the $\Gamma$-point when the spin-orbit coupling is switched on. Therefore, it is expected that 2-D Sn can host a Quantum anomalous Hall effect \cite{PhysRevB.84.195430}, similar to graphene \cite{PhysRevLett.95.226801}, silicene \cite{PhysRevB.89.035409,PhysRevLett.109.055502}, and germanene.\cite{kaloni-accept-jpcc} The magnitudes of E$_{gap}$ and $\Delta_{sp}$ become 76.3-328.7 meV and 0.71-86.3 meV, respectively, for external electric fields of 0.001-1.5 V/\AA, see Table 1. The transition from a direct to an indirect band gap is obtained for an external electric field of 2.0 V/\AA. In this case, an indirect E$_{gap}$ of 228.6 meV along the $\Gamma$-K direction and a $\Delta_{sp}$ of 90 meV at the K-point and 21 meV at the $\Gamma$-point are obtained. The external electric field results in localized eigenstates and the electronic bands at the K-point become flatter.

\begin{figure}[ht]
\includegraphics[width=0.5\textwidth,clip]{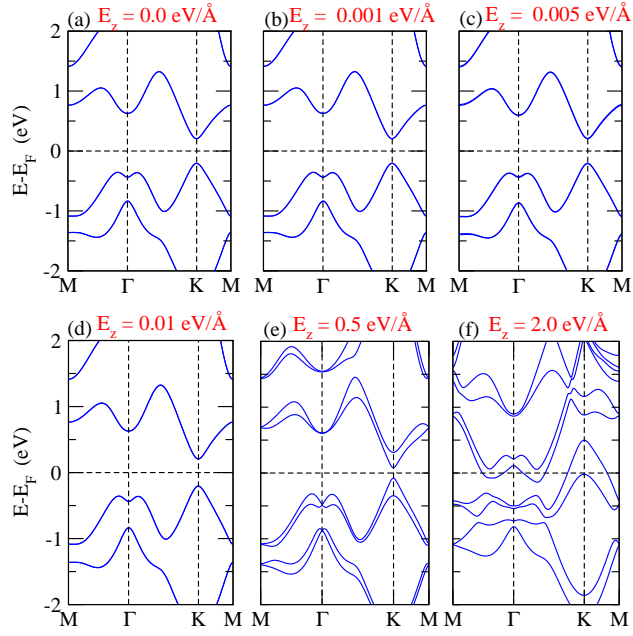}
\caption{The electronic structure of 2-D Pb for different values of the external electric field $E_z$}
\end{figure}

In Fig.\ 4, the electronic band structure of 2-D Pb is addressed. Upon inclusion of spin-orbit coupling, an E$_{gap}$ of 409.1 meV is obtained in the absence of an external electric field. In this case E$_{gap}$ is decreased with increasing field strength and the material becomes metallic for an electric field of 1.5 V/\AA\ or higher. It is worth to mention at this point that a semiconductor-to-metal transition has been demonstrated also for few layers and bulk MoS$_2$.\cite{new1} An E$_{gap}$ value of 157.2 meV with a $\Delta_{sp}$ of 233.1 meV is obtained for an external electric field of 0.5 V/\AA, see Table 1. The structural parameters agree well with a previous report.\cite{tsai2013gatedsilicene} We note that a clear trend is found for the group IV elements, especially in the nature of the band gap. A band gap of 0.0 eV, 2.0 meV, 22.9 meV, 73.4 meV, and 409.1 meV is obtained for C, Si, Ge, Sn, and Pb-based systems. The increase of the band gap with increasing atomic size can be attributed to the increase in the strength of spin-orbit coupling.   

In addition, to understand the mechanism of E$_{gap}$ and $\Delta_{sp}$, a two-band $k\cdot p$ model has been applied. This model is sufficient to describe the charge carriers in these novel 2-D materials with large intrinsic spin-orbit coupling.\cite{PhysRevB.84.195430} An effective Dirac-like Hamiltonian is used in the $xy$-plane to describe the physics of the valence and conduction bands in the $K$ and $K^{^{\prime }}$ valleys as%
\begin{equation}
\hat{H}^{\eta ,s}=v(\eta \hat{\sigma}_{x}\hat{p}_{x}+\hat{\sigma}_{y}\hat{p}%
_{y})+s\eta \lambda \hat{\sigma}_{z}+\lambda _{V}\hat{\sigma}_{z}.  \label{1}
\end{equation}%
Here, $\eta =\pm 1$ represents the $K$ and $K^{\prime }$ valleys, respectively, ($\hat{\sigma}_{x}$, $\hat{\sigma}_{y}$, $\hat{\sigma}_{z}$) is the vector of Pauli matrices (applicable to both the valence and conduction bands), $\lambda $ is the spin-orbit coupling energy, $s=\pm 1$ represents the up and down spins, respectively, and $v$ denotes the Fermi velocity of the Dirac
fermions. $\lambda _{V}=2lE_{z}$ is the term induced by the perpendicular electric field. It breaks the inversion symmetry. 
Finally, $2l$ is the perpendicular length along the buckled axis.\cite{PhysRevB.84.195430}

After diagonalization of the Hamiltonian of Eq.\ (4), the eigenvalues are obtained as%
\begin{equation}
E_{t}^{\eta,s}=t\sqrt{(v\hslash k)^{2}+(\lambda _{V}-s\eta \lambda )^{2}}
\label{2}
\end{equation}%
Here, $t=\pm 1$ represents the conduction and valence bands, respectively. Two aspects are worth noticing; (1) in the limit of zero electric field, a spin-orbit coupling induced gap ($\Delta_{sp}=\lambda$) can be obtained at wave vector $k=0$, and (2) by increasing the perpendicular electric field, both the valence and conduction bands show splitting due to a combination of spin-orbit coupling and electric field energy ($E_{gap}=\lambda_{V}-s\eta \lambda$), which for example is 22.9 meV, 73.4 meV, 409.1 meV for 2-D Ge, Sn, and Pb, respectively. With increasing perpendicular electric field, the spin-orbit splitting is enhanced and well resolved in both the bands. Moreover, all the results obtained using first-principles calculations are matched by the results from the TB approach for the lower range of the electric field, see Fig.\ 5 and Table 1. However, the TB fails to predict the properties of 2-D Pb in the presence of high external electric fields.

\begin{figure}[ht]
\includegraphics[width=0.5\textwidth,clip]{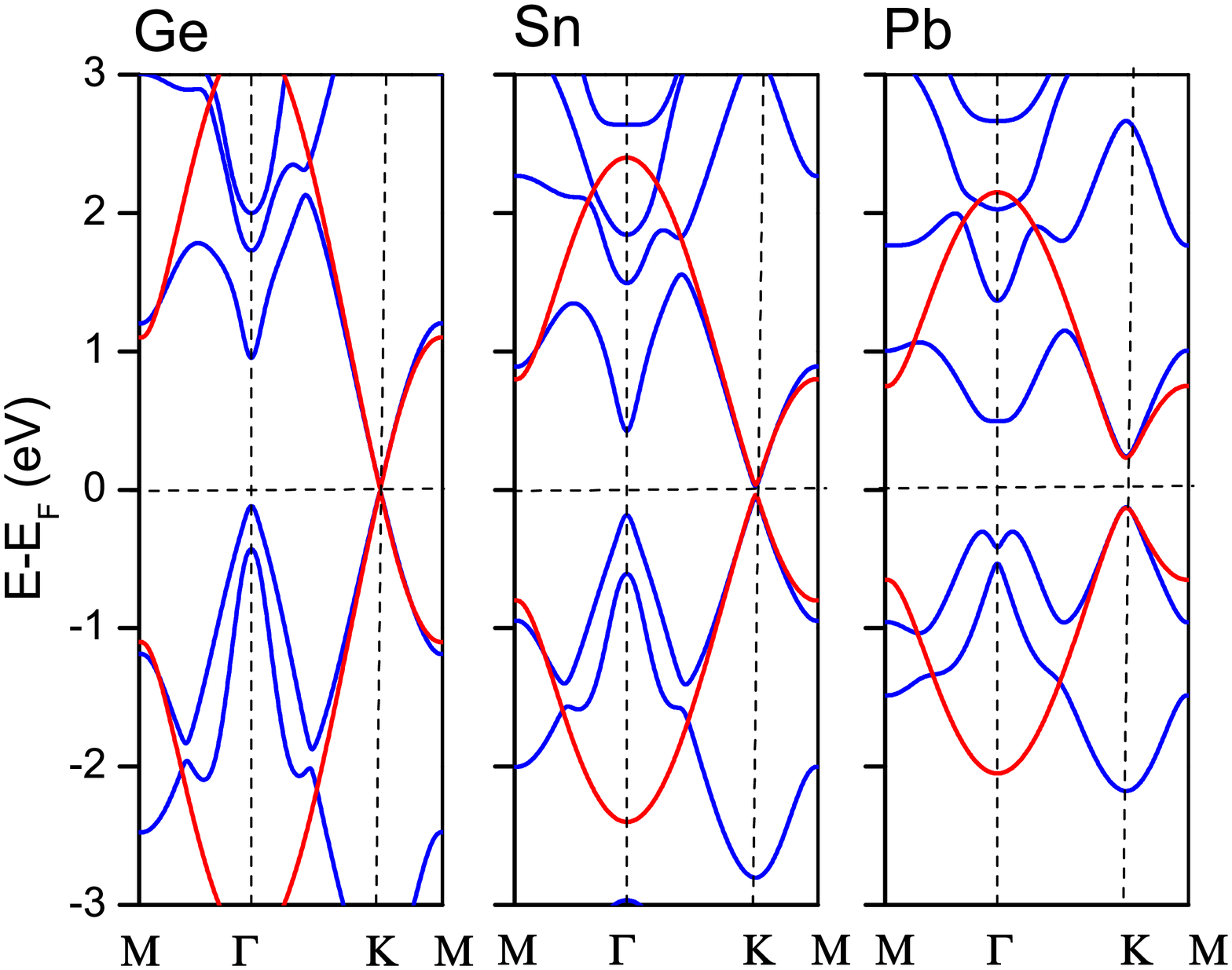}
\caption{The band structure obtained from the TB calculations (blue lines) and effective TB approach (red lines) for 2-D Ge, Sn, and Pb lattices.}
\end{figure}

\begin{table}[h]
\begin{tabular}{|c|c|c|c|c|c|c|c|c|c|}
\hline
System&$E_s$&$E_p$&$t_{ss\sigma}$&$t_{sp\sigma}$&$t_{pp\sigma}$&$t_{pp\pi}$&$\lambda$ &$t^{eff}$&$\lambda^{eff}$ \\
\hline
Ge&$-$5.65&$-$0.8&$-$1.6&2.4&2.1&$-$1.1&0.15&$-$1.1&2.5\\
\hline
Sn&$-$5.0&0.65&$-$1.2&1.9&1.7&$-$0.8&0.2&$-$0.8&7.0\\
\hline
Pb&$-$7.15&0.8&$-$0.85&1.4&1.4&$-$0.7&0.45&$-$0.7&35.0\\
\hline
\end{tabular}
\caption{The TB parameters for 2-D Ge, Sn, and Pb lattices, where $E_s$ and $E_p$ are the on-site energy for \textit{s} and \textit{p} orbitals and all the parameters are measured in eV except $\lambda^{eff}$, which is measured in meV.} 
\end{table}

The effective TB model describes the band structure at the K-point effectively but it fails around the $\Gamma$-point. In order to fit the strength of the spin-orbit interaction, the size of the electronic gap and band splitting at the K-point are considered. In Table 2, the TB parameters for 2-D Ge, Sn, and Pb are summarized. The TB parameters are obtained according to the method of Ref.\ \cite{PhysRevB.78.205425} $E_{\alpha}$ is the on-site energy and $t_{\alpha\beta}$ are different components of the hopping between nearest neighbor sites. The obtained $\lambda$ is proportional to the atomic mass of each element. By increasing the atomic radius the effective hopping and spin-orbit strength decreases and increases, respectively. The TB parameters obtained in this study could be useful for future calculations of nanostructures as will be illustrated in the following section.

\subsection{TRANSPORT PROPERTIES}
\begin{figure}[ht]
\includegraphics[width=0.5\textwidth,clip]{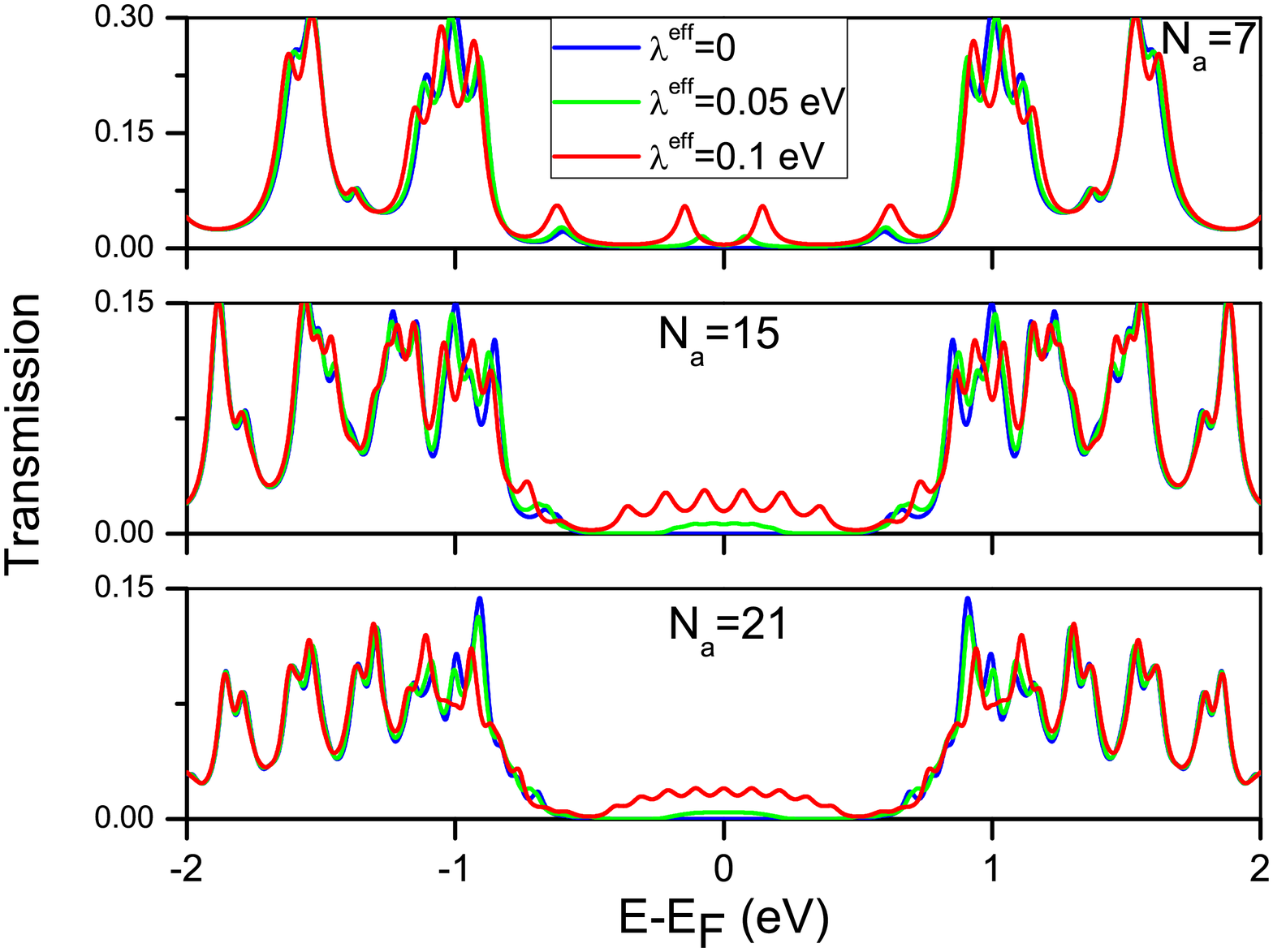}
\caption{The transmission spectrum of TQD between two electrodes for N$_a =$ 7, 15, and 21 atoms.}
\end{figure}

In the following the transport properties are studied for TQD by using the effective tight binding model. Triangular shaped graphene quantum dots have been synthesized experimentally \cite{nl300897m} and it has been suggested that quantum dots made from the other group IV element should be triangular in shape as well. The ratio $\lambda^{eff}$/$t^{eff}$ varies from zero for graphene up to 0.05 for Pb. For the self-energy, only the hopping between electrode and the edge atoms of the TQD is considered and the imaginary part of the self-energy is set to 0.05 eV. The TQD is characterized by the number of atoms on each edge N$_a$. The transmission spectrum of different size quantum dots is addressed in Fig.\ 6.

The transmission is symmetric and peaks are corresponding to eigenstates of the Hamiltonian. By increasing the size of the quantum dot, the length of the dot increases, and the transmission through the nanostructure decreases. In the absence of spin-orbit coupling the triangular dot has (N$_a-3$)/2 degenerate and localized wave functions in the zero state energy that do not contribute to the transmission. The spin-orbit coupling removes the degeneracy and results in localization of the wave functions; the transmission gap is filled with peaks. For larger dots the energy difference between states is decreased and the peaks in the gap become flat. The first two peaks around $-1$ eV in the transmission curve are corresponding to the first electronic states. By increasing the strength of spin-orbit coupling the first transmission peaks are decreased. The applied voltage difference between left and right leads to an opening of the energy window and current flows in the quantum dot. The transmission peaks in the gap lead to steps for low voltages in the $I-V$ curves. By applying an external electric field perpendicular to the dot surface the energies of the localized states are shifted to lower or higher energies depending on the direction of the electric field. As a result the transmission spectrum becomes asymmetric with respect to the Fermi level.

\section{CONCLUSION}
In summary, three different 2-D hexagonal lattices, specifically Ge, Sn, and Pb, were investigated using first-principles (DFT) and tight-binding approaches. The electronic structure of 2-D Ge, Sn, and Pb without and with the presence of a perpendicular electric field in combination with spin-orbit coupling have been addressed. The obtained results are compared with the tight-binding approach. The nature of the band gap, a transition from semiconductor to metal, a transition from direct to indirect band gap, and band splitting are found and discussed. A large splitting and tuning of the band gap are obtained by varying the perpendicular external electric field, which is expected to be useful for spintronic device applications. We emphasize that silicene-based transistors have been realized experimentally already, and they can be operated at room temperature \cite{natnatech2015}. Thus we are also expecting interesting results for our systems under study. In addition, the $k\cdot p$ model is considered in order to understand the band splitting. Excellent agreement between the first-principles and tight-binding approaches is found. Furthermore, the transport properties of typical 0-D triangular quantum dots between two electrodes are addressed, which will pave the way to the experimental realization \cite{nl300897m}.

\section{Acknowledgements}
GS acknowledges funding from the Natural Sciences and Engineering Council of Canada (NSERC, Discovery Grant). 
MSF acknowledges support by the Natural Sciences and Engineering Research Council (NSERC) of Canada, the 
Canada Research Chair program, Canada Foundation for Innovation (CFI), the Manitoba Research and Innovation Fund, 
and the University of Manitoba.

\bibliography{revised_manuscript_GS_Mar24}
\bibliographystyle{apsrev4-1}

\end{document}